\documentclass[notitlepage,twocolumn,prl,showpacs,floatfix]{revtex4-2} 

\usepackage{amsmath,amssymb,amsfonts,latexsym,bm}
\usepackage{graphicx,epsfig}

\usepackage[svgnames]{xcolor}
\usepackage{xfrac}
\usepackage{mathrsfs}

\usepackage[colorlinks,linkcolor=MediumBlue,citecolor=MediumBlue,urlcolor = MediumBlue]{hyperref}
\usepackage{url}

\allowdisplaybreaks

\newcommand{\Pe}{{\rm Pe}}

\begin{document}

\noindent \textbf{Comment on ``Flocking without Alignment Interactions in Attractive Active Brownian Particles''}\\

It was recently argued in~\cite{Caprini2023PRL} that attractive active Brownian particles (ABPs) may self-organize in two dimensions into a flocking state 
characterized by the formation of a cluster with nonzero global order of velocity orientations~\footnote{The particle velocities include both contributions from the self-propulsion and pairwise interactions. 
Due to the absence of aligning interactions in the dynamics, the self-propulsion orientations alone are always disordered.} and scale-free correlations.
Increasing the self-propulsion strength of particles --measured by the dimensionless Péclet number $\Pe$--
leads to the break-up of the cluster into a globally disordered state.
While the phenomenology reported in~\cite{Caprini2023PRL} may seemingly resemble flocking, 
a closer analysis of finite size effects reveals that it lacks several of its defining characteristics such as long-range correlations and large-scale directed motion.

\begin{figure}[t!]
    \includegraphics[width=.99\columnwidth]{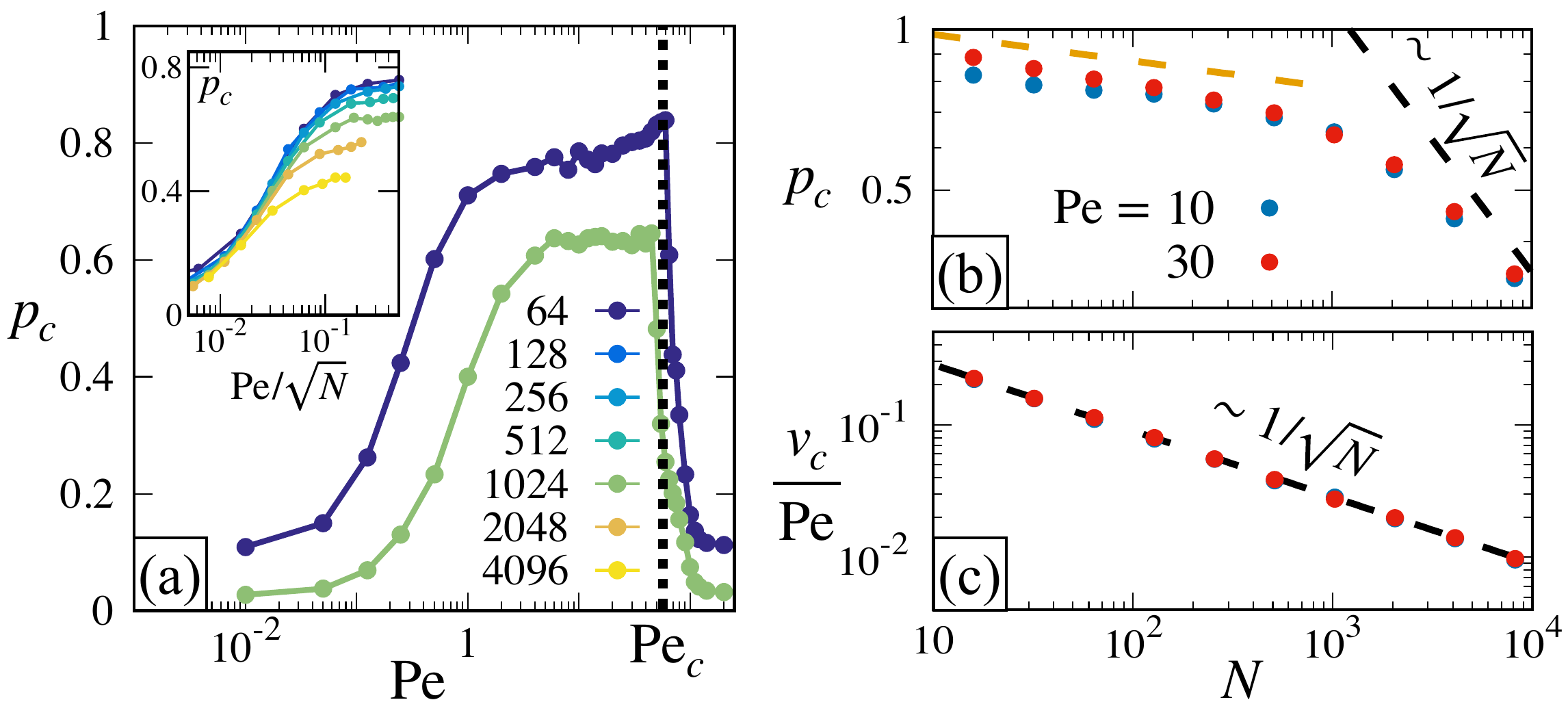}
    \caption{(a) Velocity polarization as function of the Péclet number for various cluster sizes.
    Inset: $p_c$ vs.\ $\Pe/\sqrt{N}$ for small $\Pe$ values.   
    (b,c) $p_c$(b) and mean particle velocity rescaled by $\Pe$(c) as function of $N$ with fixed $\Pe$. 
    In (b) the orange dashed line shows a power law decay with exponent $-0.05$ as guide for the eyes.
    In (c) the two sets of data overlap.
    Unspecified simulation parameters are identical to those used in~\cite{Caprini2023PRL}.}
    \label{fig1}
\end{figure}

The ordering transition described in~\cite{Caprini2023PRL} arises when {\it reducing} the strength of activity ($\Pe$),
whereas flocking must ultimately disappear in the equilibrium limit $\Pe = 0$.
Simulations similar to those of~\cite{Caprini2023PRL} indeed reveal the formation of a crystalline cluster containing all particles arranged on a hexagonal lattice. 
Figure~\ref{fig1}(a) shows that the velocity polarization $p_c = N^{-1}\langle | \sum_{i=1}^N \bm v_i / |\bm v_i| | \rangle$, where the $\bm v_i$ are the particle velocities and $N$ 
is their total number, takes appreciable values only for $\Pe^* < \Pe < \Pe_c$.
Importantly, $\Pe^* \sim \sqrt{N}$ (inset of Fig.~\ref{fig1}(a)) while
the upper threshold $\Pe_c$ beyond which the cluster dissolves varies little with $N$ (Fig.~\ref{fig1}(a) and~\cite{Caprini2023PRL}).
Since $\Pe = \ell_p/\sigma$, where $\ell_p$ and $\sigma$ are respectively given by the persistence length of the active motion and inter-particle distance in the crystal, 
$\Pe/\sqrt{N}$ essentially compares the persistence length with the size of the cluster.
The scaling $\Pe^* \sim \sqrt{N}$ thus implies that the velocity correlations of ABPs appear scale-free on scales below $\xi \propto \ell_p$, which corresponds to the regime studied in~\cite{Caprini2023PRL}, 
but decay exponentially beyond this finite correlation length.
Hence, the polarization always disappears in large enough systems.

The finite size dynamics of the cluster also exhibits a number of properties that set it apart from flocking.
Although $p_c$ may take large values at low $N$, the corresponding time series are strongly intermittent (not shown). 
The decay of $p_c$ with $N$ at fixed $\Pe$ shown in Fig.~\ref{fig1}(b) reveals a crossover between two power laws. 
At large $N$, the corresponding scaling exponent approaches $-\tfrac{1}{2}$, which confirms that the velocity orientations are asymptotically disordered.
In turn, the negative exponent $> -\tfrac{1}{2}$ found at smaller sizes rather points to an equilibrium-like quasi-long-range order.
This situation is at odds with most flocking systems which are truly ordered even in two dimensions~\cite{DADAM}.  
Another hallmark of flocking is the coupling between the order parameter and the particle velocities~\cite{DADAM}, leading to globally directed motion.
Figure~\ref{fig1}(c) shows that, in fact, the mean velocity $v_c \equiv N^{-1} \langle | \sum_{i=1}^N \bm v_i |\rangle$
self-averages to zero even in the regime where the polarizations are quasi-ordered.
Therefore, the global velocity of the cluster is statistically equivalent to that of noninteracting ABPs. 

Dense suspensions, whether glassy or liquid-like, of purely repulsive ABPs also see their velocities correlated over scales $\propto \ell_p$~\cite{HenkesNatComm2020,Szamel_2021,Bettolo_Marconi_2021}.
Although further studies are required to pinpoint the relative contributions leading to this behavior in models of attractive ABPs, 
the apparent similarities between the two cases suggest that the role of attraction for its emergence is mainly restricted to keeping the cluster cohesive.
At any rate, studying how velocity correlations may couple to or influence other types of order in active crystals~\cite{KlamserNatCom2018,Digregorio2018PRL,Paliwal2020PRR,HuangPRE2021,ShiPRL2023} provides an interesting research avenue.
\\

\noindent Benoît Mahault\\
\indent {\small Max Planck Institute for Dynamics and Self-Organization (MPI-DS), 37077 G\"ottingen, Germany}\\

\acknowledgements
I thank Hugues Chat\'e, Yu Duan, Ramin Golestanian, and Andrej Vilfan for interesting discussions and helpful comments on the manuscript.

\bibliographystyle{apsrev4-2}
\bibliography{biblio}

\end{document}